\documentclass[prc,twocolumn,showpacs,nofootinbib,preprintnumbers,unsortedaddress,amsmath,amssymb,floatfix]{revtex4}
\usepackage{ulem}
\usepackage{color}
\usepackage{graphicx}
\usepackage{bm}

\newcommand{\be}{\begin{equation}}
\newcommand{\ee}{\end{equation}}
\newcommand{\bea}{\begin{eqnarray}}
\newcommand{\eea}{\end{eqnarray}}

\begin{document}

\title{Quenching factor of Gamow-Teller and spin dipole giant resonances}

\author{Li-Gang Cao$^{1}$, Shi-Sheng Zhang$^{2}$, H. Sagawa$^{3,4}$
}

\affiliation{
$^{1}$School of Mathematics and Physics, North China Electric Power University, Beijing 102206, China\\
$^{2}$School of Physics and Nuclear Energy Engineering, Beihang University, Beijing 100191, China\\
$^{3}$RIKEN, Nishina Center, Wako 351-0198, Japan\\
$^{4}$Center for Mathematics and Physics, University of Aizu, Aizu-Wakamatsu, Fukushima 965-8580, Japan\\
}
\date{\today}

\begin{abstract}
Gamow-Teller (GT) and spin-dipole (SD) strength distributions
of four doubly magic nuclei $^{48}$Ca, $^{90}$Zr, $^{132}$Sn  and $^{208}$Pb
are studied by the self-consistent  Hartree-Fock plus random phase approximation (RPA) method. The Skyrme forces SAMi and SAMi-T without/with
tensor interactions are adopted in our calculations. The calculated strengths are compared with available experimental data.
The RPA results of GT and SD strengths of all four nuclei show  fine agreement with observed GT and SD resonances in energy.
A small GT peak below the main GT resonance is better described by the Skyrme interaction SAMi-T with the tensor terms.
The  quenching factors for GT and SD are extracted from the comparisons between RPA results and  experimental  strengths.
It is pointed out that the quenching effect on experimental SD peaks is somewhat modest compared with that on GT peaks in the four nuclei.
\end{abstract}

\pacs{21.60.Jz, 24.30.Cz, 25.40.Kv, 21.60.Ev}

\maketitle

\section{Introduction}
Spin-isospin excitations provide a unique opportunity to
study the spin correlations in nuclei \cite{1}.
Among them, the Gamow-Teller (GT) transition is
the simplest with   both spin and isospin transfers by one unit,
 but no transfer of other
quantum numbers. The next one is the spin-dipole (SD) excitations which involve the orbital angular momentum transfer by one unit together with the spin and isospin transfer. At small momentum
transfers, the spin-isospin particle-hole interaction is strongly
repulsive, and the residual interaction leads to collective
excitations such as the GT and SD resonances. The quenching of the total GT strength \cite{2}
from the model-independent  sum rule \cite{3} (also called the Ikeda sum rule \cite{4})
 has prompted theoretical studies of possible
mechanisms, ranging from conventional configuration mixing
\cite{5,6} to an admixture of the $\Delta-$hole (-h) states \cite{7,8,9,10}.
Experimental investigations into the $(p, n)$ \cite{11} and $(n, p)$ \cite{12}
reactions of $^{90}$Zr using the multipole decomposition (MD)
technique \cite{13} have revealed that configuration mixing effects,
such as coupling to 2-particle2-hole (2p-2h) excitations, play
an important role in GT quenching, whereas $\Delta-h$ coupling has
a minor role. It has also been noted that some quenching may
result from tensor interaction effects that couple the GT states
with the spin-quadrupole 1$^+$ states \cite{14}.

 MD
analysis of $(p, n)$ cross sections has identified a considerable
amount of broadly distributed $L=1$ strength at excitation energies beyond the main GT peak \cite{15}.
This $L=1$ strength is nothing but the spin-dipole strength.
 The spin-dipole components were extracted from  $^{90}$Zr$(p, n)$
and $^{90}$Zr$(n, p)$ data by MD analysis
 assuming a proportionality relation
between the SD cross section and the relevant transition strength \cite{16,17}.
   It should
be noted that the experimental strengths include all the SD
strengths with spin-parity transfer $J^{\pi}=0^-, 1^-$, and 2$^-$
because the separation of the individual multipole contributions is
difficult in the MD analysis \cite{13}.   The spin analyzing power measurements were performed with the polarized protons for the charge exchange reaction $^{208}$Pb$(p,n)^{208}$Bi, and each multipole component is successfully separated from the total strength \cite{Wakasa12}.
The separated SD strengths
should be useful for further theoretical investigations on
the tensor interaction effects on SD excitations
\cite{19,20,21,22},  and also the neutron matter equation of state \cite{23}.

In the astrophysical context,  the spin-isospin mode get much attraction these days;
 $\beta$-decay probabilities have essential roles for the $r$-process nucleosynthesis together with nuclear masses, and photonuclear
cross sections \cite{Kajino2019}.
 We should mention also the importance of knowing the
neutrino-nucleus interactions with axial-vector currents in  the
stellar environment \cite{neutrino-A}.
  All these problems motivate  the recent works concerning the spin-isospin
nuclear modes and the quenching of axial-vector currents.

Double beta decay processes have been getting much attention recently to study the neutrino mass problem, which is predicted by beyond the standard model of elementary particles.  Two types of double beta decay have been discussed. One is 2-neutrino (2$\nu$) double beta decay and another one is 0$\nu$ double beta decay. The latter process is held with Majorana neutrino. The 2$\nu$ double beta will occur through GT states at the intermediate states, while 0$\nu$ double beta decay goes through spin-isospin excitations of any angular momentum $l=0, 1, 2,\cdot\cdot\cdot$,  in which GT and SD may play the most important role. The quenching factors of these spin excitations are quite important for quantitative predictions of these double beta decays. There have been many discussions of GT quenching with respect to the GT Ikeda sum rules \cite{11,12,13}.  On the other hand,  the quenching of SD states have not  much discussed so far.

In this paper, we study GT and SD states in  four  doubly magic nuclei $^{48}$Ca, $^{90}$Zr, $^{132}$Sn  and $^{208}$Pb by using
the self-consistent Hartree-Fock (HF) plus random phase approximation (RPA) model with/without tensor interactions.  We adopt modern energy density functions (EDFs) SAMi \cite{SAMi} and SAMi-T  \cite{SAMi-T}
for the theoretical study.   The latter has tensor terms which are determined from "ab initio" type Bruckner HF calculations with AV18 interaction.  The paper is organized as follows.
Section II is devoted to theoretical models for the HF+RPA
calculations.  Results are given in Sec. III in comparisons with experiments.
A smmary is given in Sec. IV.

\section{Sum rules and theoretical models}
The GT transition operator is defined as
\be \label{GT-ope}
\hat{O}_{\pm}=\sum_{i=1}^A\sum_\mu\sigma_\mu(i) t_\pm(i)
\ee
where $\sigma_\mu$ is the spin operator and $t_\pm=t_x\pm it_y$ are the isospin raising and lowering operators, respectively.
The model-independent sum rule can be evaluated for the operators \eqref{GT-ope} as
\be \label{GT-sum}
S_--S_+=\sum_f|\langle f|\hat{O}_{-}|i\rangle|^2-\sum_f|\langle f|\hat{O}_{+}|i\rangle|^2 =3(N-Z)
\ee
where $|i\rangle$ and $|f\rangle$ are the initial and final states excited by the Gamow-Teller operator, respectively, and
$N$ and $Z$ are neutron and proton numbers, respectively.  This sum rule is often referred to as the "Ikeda" GT sum rule.

The study of the charge-exchange spin-dipole (SD)
excitations of $^{208}$Pb (inspired by recent accurate measurements
\cite{Wakasa12})
and of $^{90}$Zr will be shown to elucidate in a quite specific way
the effect of tensor correlations.
To get an unambiguous signature of the effect of the tensor
force, which is strongly spin-dependent,
one can expect that the separation of the
strength distributions of the $\lambda^\pi=0^-$, $1^-$ and $2^-$ components
is of great relevance.
The charge-exchange SD operator is defined as
\begin{eqnarray}
\hat{O}_{\pm
}^{\lambda}=\sum\limits_{i}\sum_{\mu}t_{\pm
}(i)r_{i} \left[ Y_{l=1}(\hat r_i)
\otimes \sigma(i) \right]^{(\lambda\mu)}, \label{multipole}
\end{eqnarray}
where $Y_{lm}$ is the spherical harmonics.
The $n-$th energy weighted sum rules $m_n$ for the
$\lambda$-pole SD operator are defined as
\begin{eqnarray}
m_n^{\lambda}(t_{\pm})=\sum\limits_{f}(E_f-E_i)^n|\langle f|
\hat{O}_{\pm}^\lambda|i \rangle|^2,
\end{eqnarray}
and
the sum rule which is known to hold is
\be \label{eq:SD-sum}
m_0^{\lambda}(t_{-})
-m_0^{\lambda}(t_{+})=
\frac{2\lambda+1}{4\pi} (N \langle r^2 \rangle_n-Z \langle r^2
\rangle_p) ,
\ee
where $\langle r^2 \rangle_n ( \langle r^2
\rangle_p)$ is the mean square radius of neutrons (protons).
This SD sum rule is analytically exact, but depends on the neutron skin size which has some variations in microscopic models.

 In this section, we will briefly report the theoretical method of
our calculations. More detailed information about the Skyrme HF plus RPA
  calculations can be found in Ref. \cite{Colo12,Frac05}.
First, we start by solving the Skyrme HF equation in the coordinate
space, the radial mesh is 0.1 fm for $^{48}$Ca ($^{90}$Zr, $^{132}$Sn, $^{208}$Pb),
and the maximum value of the radial coordinate is set to be 20 fm for $^{48}$Ca ($^{90}$Zr, $^{132}$Sn, $^{208}$Pb), respectively.
   In order to calculate
unoccupied states at positive energy, the continuum has been
discretized by adopting the box boundary condition.
Thus, we get
the energies as well as the wave functions for particle (p) and hole
(h), which are the input for RPA 
 calculations. We solve the
RPA equations in the  matrix formulation; all the hole states
are considered when we build particle-hole (p-h) configurations,
while for the particle states we choose the lowest ten
unoccupied states for each value of $l$ and $j$.
 We adopt SAMi and SAMi-T as EDFs for numerical calculations. SAMi EDF is designed for good description of spin-isospin mode \cite{SAMi}.  The EDF, SAMi-T, has tensor terms, which are determined by "ab initio" type Bruckner HF results with AV18 interaction \cite{SAMi-T}.

\section{Results}
We calculate GT states and SD states in the self-consistent HF+RPA model.  The GT states are studied to validate the model predictions for the peak energies and also  to check the effect of tensor interactions.
We will study also how much the sum rule values of GT and SD strength distributions are affected  by the tensor interactions.
\subsection{GT states}
\begin{figure}[hbt]
\includegraphics[width=0.45\textwidth]{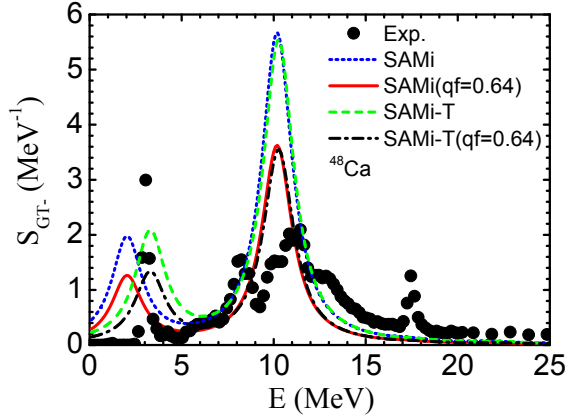}
\vglue -2.5cm
\caption{RPA strength functions of $^{48}$Ca for $t_-$ channel of GT resonance. The solid circles are the experimental data taken from ref. \cite{Yako2009}. The short-dotted (short-dashed) and solid (dashed-dotted) lines are the theoretical results without and with a quenching factor given by using SAMi (SAMi-T) EDF.} \label{GT-Ca}
\end{figure}

\begin{figure}[hbt]
\includegraphics[width=0.45\textwidth]{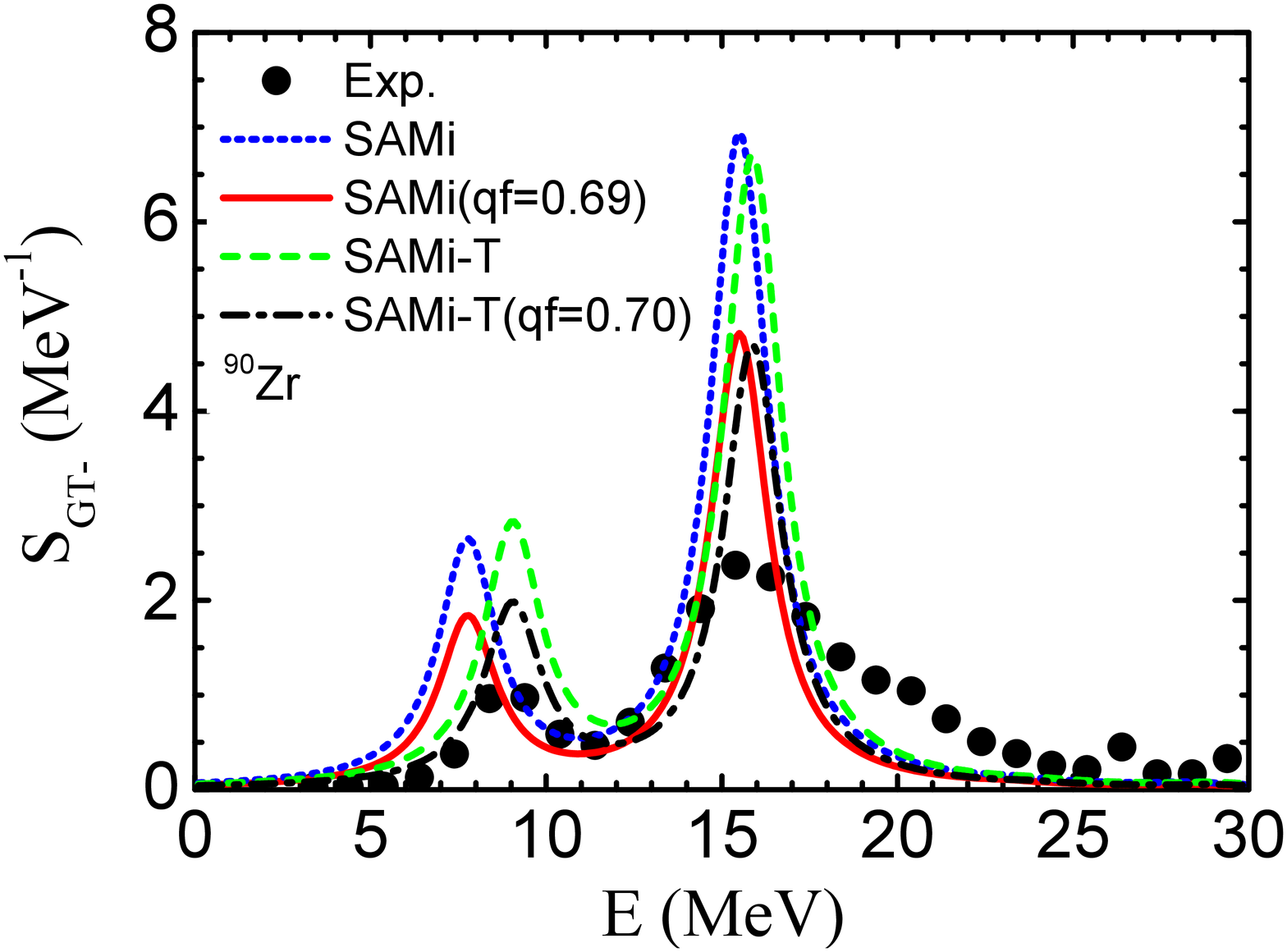}
\vglue -2.5cm
\caption{The same as for Fig. \ref{GT-Ca}, but for $^{90}$Zr. Experimental data are taken from ref. \cite{11}
}
\label{GT-Zr}
\end{figure}

\begin{figure}[hbt]
\includegraphics[width=0.45\textwidth]{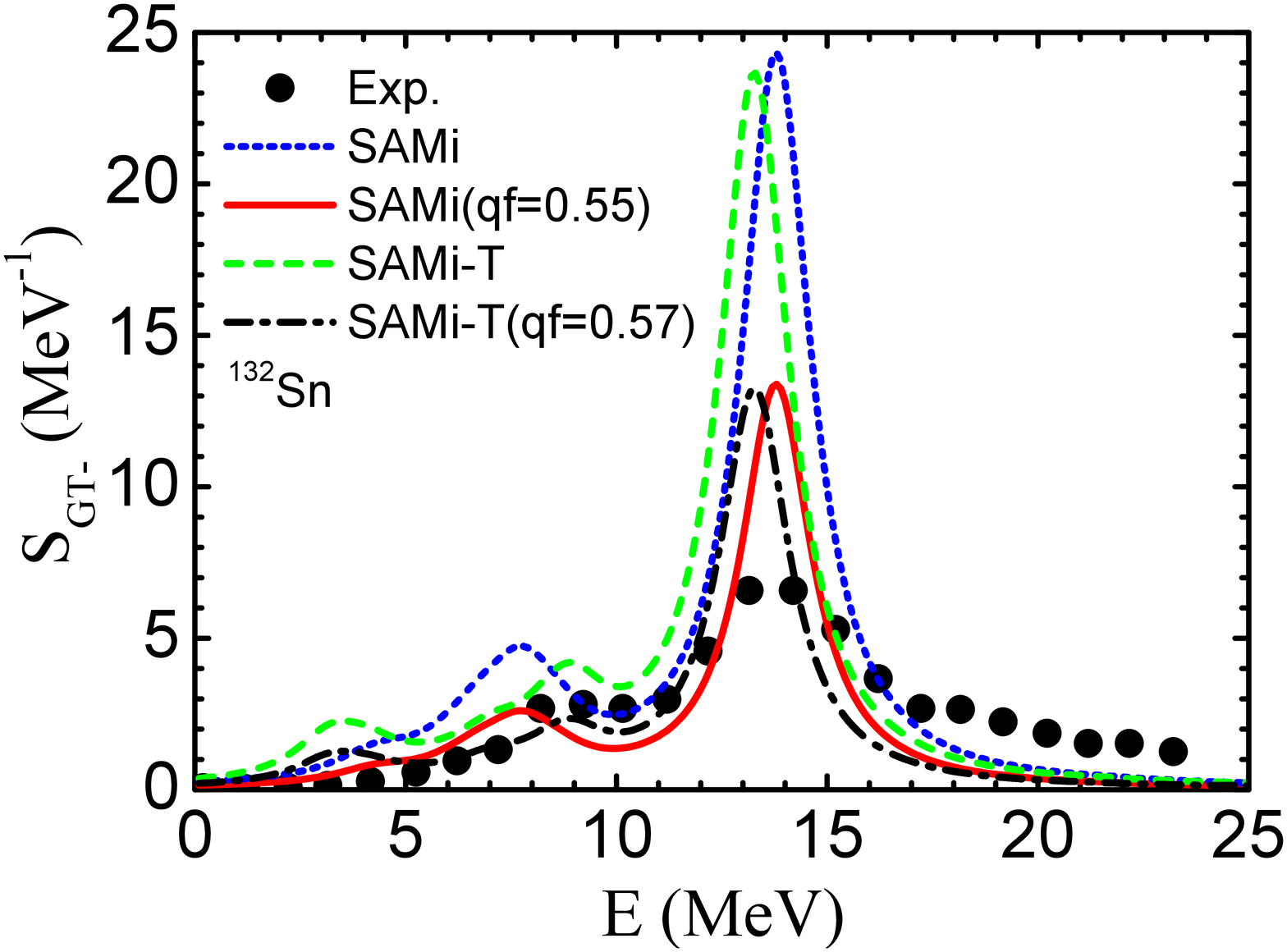}
\vglue -2.5cm
\caption{The same as for Fig. \ref{GT-Ca}, but for $^{132}$Sn.  Experimental data are taken from ref. \cite{Yasuda2018}
} \label{GT-Sn}
\end{figure}

\begin{figure}[hbt]
\includegraphics[width=0.45\textwidth]{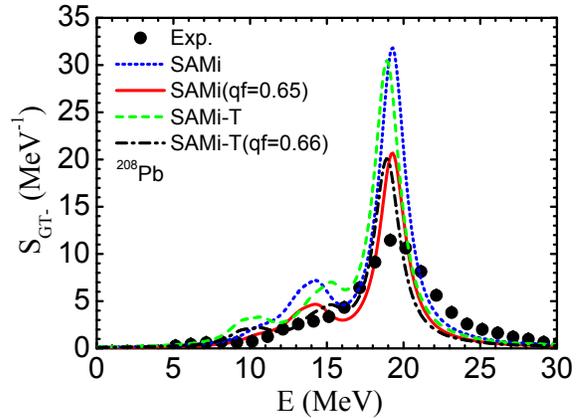}
\vglue -2.5cm
\caption{The same as for Fig. \ref{GT-Ca}, but for
$^{208}$Pb.  Experimental data are taken from ref. \cite{Wakasa12}
} \label{GT-Pb}
\end{figure}

GT results are shown in Figs. \ref{GT-Ca}, 
   \ref{GT-Zr}, \ref{GT-Sn}, \ref{GT-Pb}  for $t_-$ 
    channel  of $^{48}$Ca,  
      $^{90}$Zr, $^{132}$Sn and $^{208}$Pb, respectively.
The main experimental  GT resonance was found experimentally at E$_x\sim$10MeV and also a small peak at E$_x$=3MeV in
$^{48}$Ca.  The calculated results with SAMi reproduces well the main peak, but predicts 1MeV lower than the experimental one for the low-energy strength.  The SAMi-T gives essentially the same results for the main peak, but give a better excitation energy in comparison with the experimental one.
The integrated GT strength from E$_x=0\to25$MeV is 15.3, which is 64\% of the GT sum rule in Eq. \eqref{GT-sum}.  The calculated results exhaust almost 100\% of the sum rule up to E$_x$=20MeV.   The quenching factor for the transition strength is defined as
\be
qf=\frac{\sum_{E_x=0}^{E_x(max)}B(GT:E_x)_{exp}}{\sum_{E_x=0}^{E_x(max)}B(GT)_{cal}}
\ee
where $E_x(max)$ is taken to be 25MeV for $^{132}$Sn and to be 30MeV for $^{48}$Ca, $^{90}$Zr and $^{208}$Pb in the GT case.  The quenching factor $qf=0.64$ corresponds to
a renormalization factor of $q_{RF}=0.80$ for the GT  transition operator to retain the empirical sum rule value.

\begin{figure}[hbt]
\includegraphics[width=0.5\textwidth]{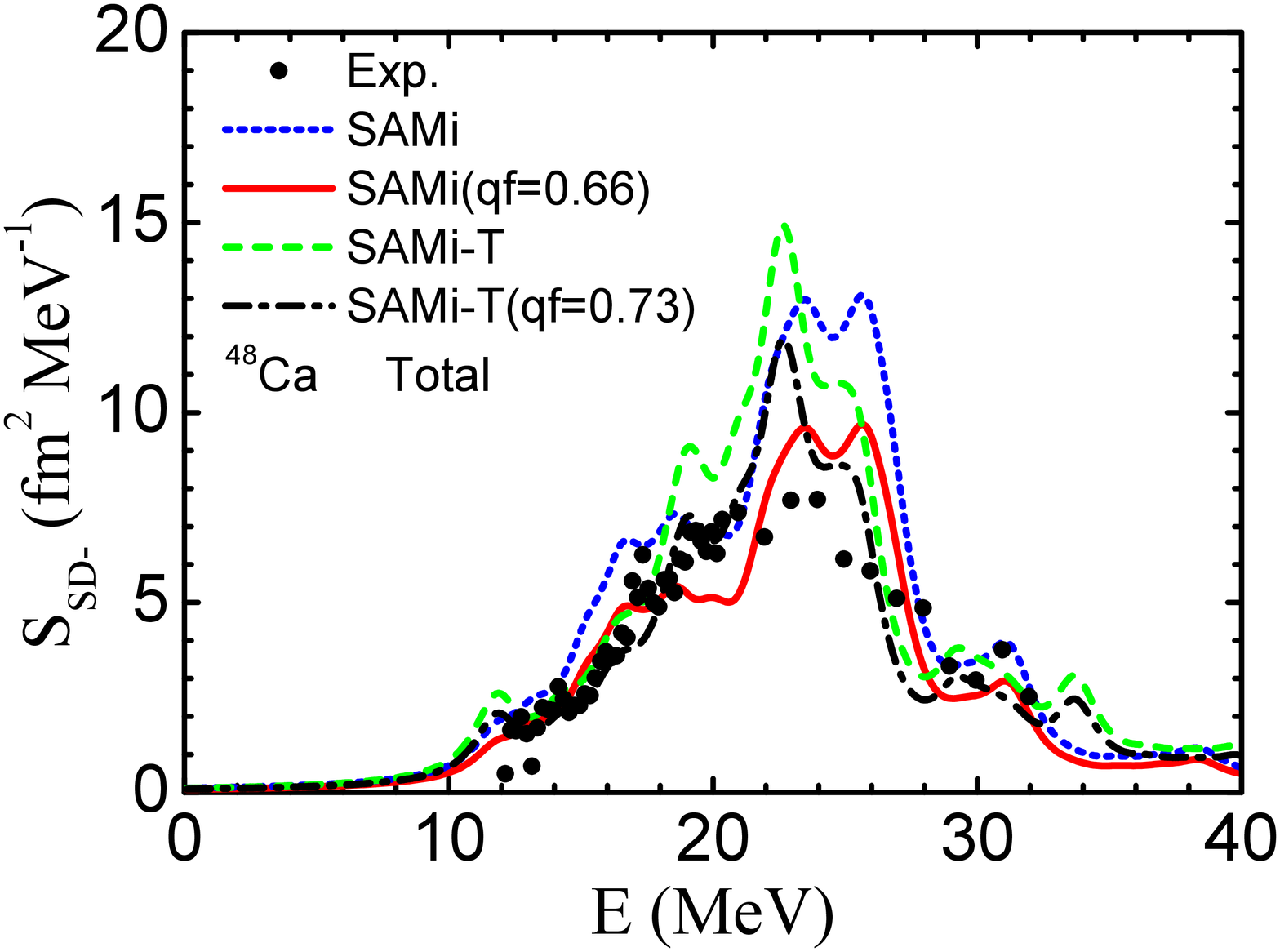}
\vglue -2.5cm
\caption{RPA strength functions of $^{48}$Ca for SD$_-$ resonance. The short-dotted (short-dashed) and solid (dashed-dotted) lines are the theoretical results without and with a quenching factor calculated by using SAMi (SAMi-T) EDF.
Experimental data are taken from ref. \cite{Yako2009}.} \label{SD-Ca}
\end{figure}

\begin{figure}[hbt]
\includegraphics[width=0.5\textwidth]{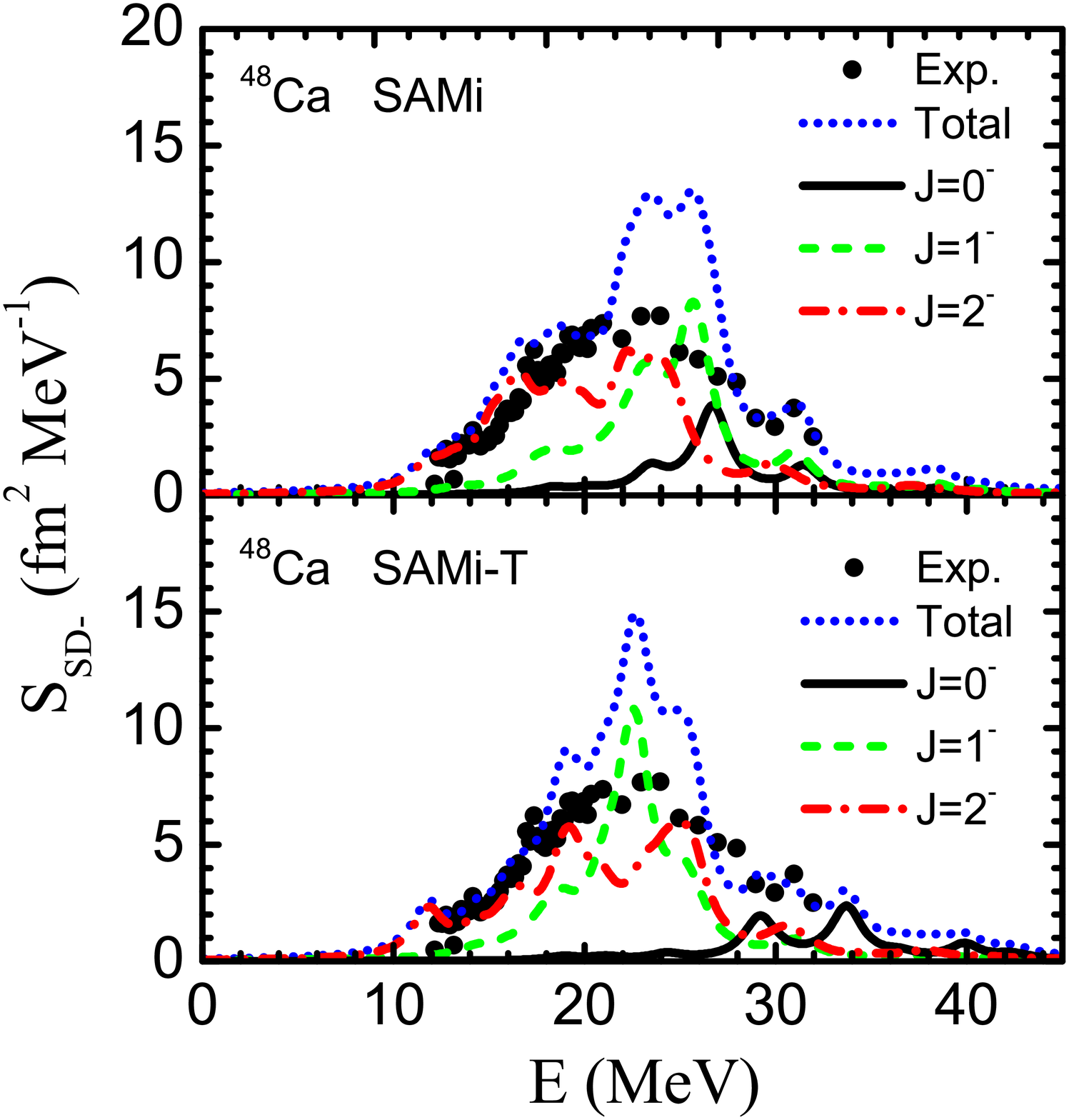}
\caption{RPA strength functions of $^{48}$Ca for SD$_-$ resonance calculated by using SAMi (SAMi-T) EDF. The total strength and the J$^{\pi}$= 0$^-$, 1$^-$, and 2$^-$ components are shown. Experimental data are taken from ref. \cite{Yako2009}.} \label{SD-Ca-J}
\end{figure}

\begin{figure}[hbt]
\includegraphics[width=0.5\textwidth]{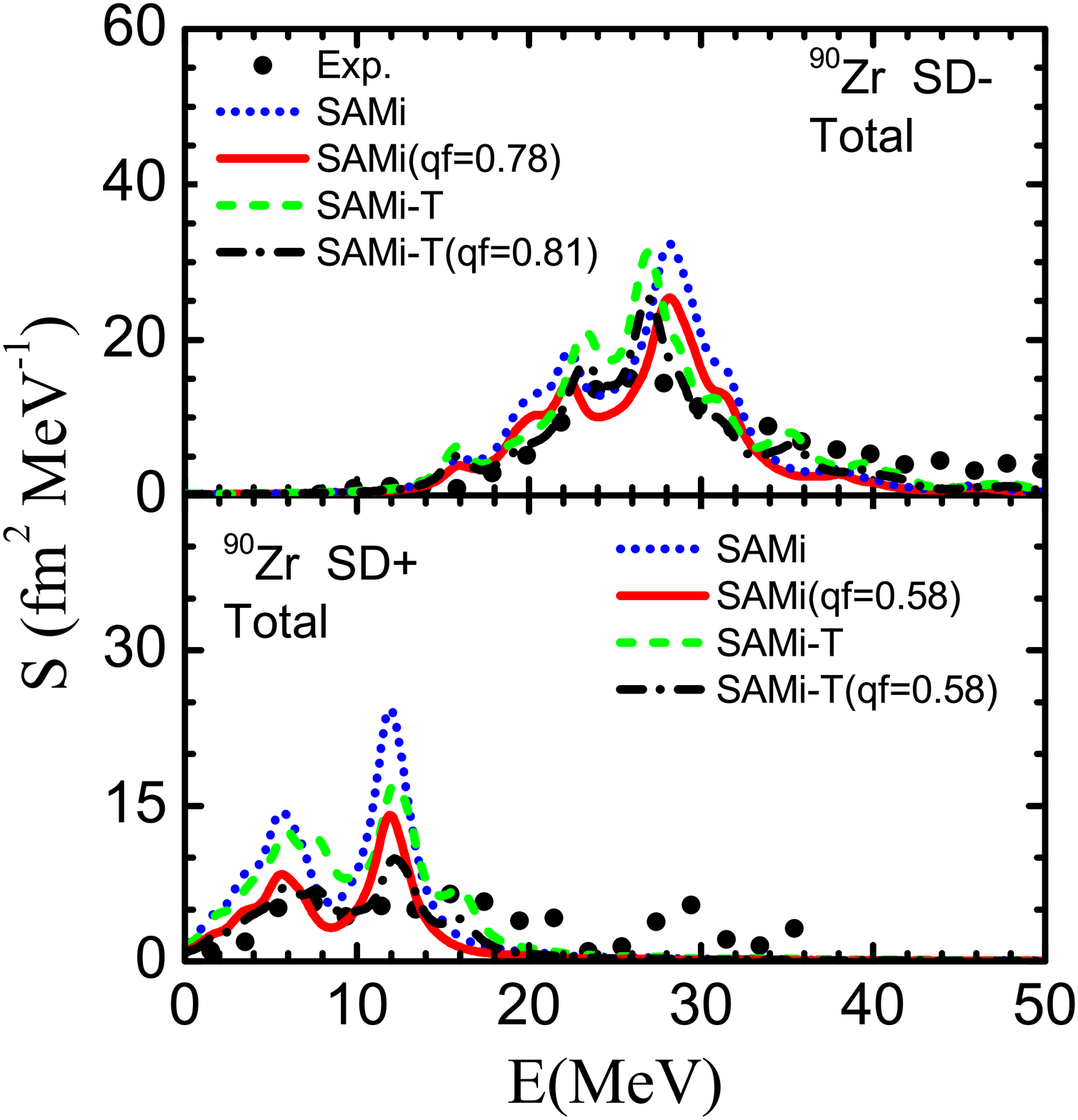}
\caption{RPA strength functions of $^{90}$Zr for SD$_-$ and SD$_+$ resonances. The short-dotted (short-dashed) and solid (dashed-dotted) lines are the theoretical results without and with a quenching factor calculated  by using SAMi (SAMi-T) EDF. Experimental data
 are taken from ref. \cite{16}.} \label{SD-Zr}
\end{figure}

\begin{figure}[hbt]
\includegraphics[width=0.5\textwidth]{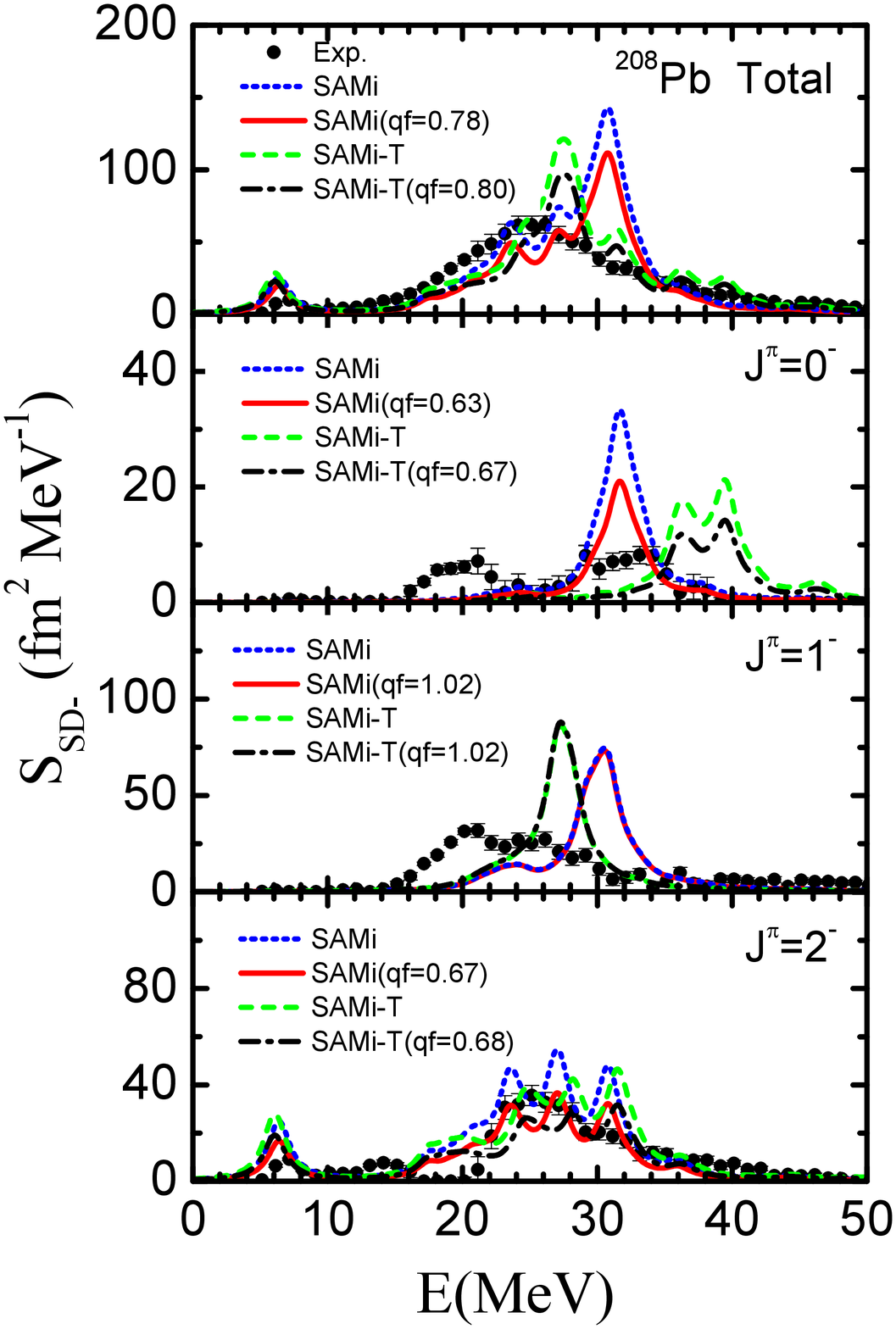}
\vglue -0.5cm
\caption{RPA strength functions of $^{208}$Pb for SD$_-$ resonance, from top to bottom, the total strength and the J$^{\pi}$= 0$^-$, 1$^-$, and 2$^-$ components are shown. The short-dotted (short-dashed) and solid (dashed-dotted) lines are the theoretical results without and with a quenching factor calculated by using SAMi (SAMi-T) EDF.  Experimental data are taken from ref. \cite{Wakasa12}.} \label{SD-Pb}
\end{figure}

The calculated GT strength in $^{90}$Zr with SAMi gives a good account of empirical GT peak at E$_x\sim$15.5MeV.  The SAMi-T EDF
gives essentially the same response for the GT peak. However, the low energy peak is better described by SAMi-T than SAMi.  The quenching factors for the two calculations are almost the same as $qf=0.69$ and $qf=0.70$ for SAMi and SAMi-T, respectively.
For $^{132}$Sn, the main peak is shifted 1MeV lower by SAMi-T than SAMi, while the small shoulder at 8MeV by SAMi is shifted to 1MeV high by SAMi-T.  The quenching factor for SAMi and SAMi-T are $qf=0.55$ and $qf=0.57$, respectively.   For $^{208}$Pb in Fig. \ref{GT-Pb}, the general trend is the same as that of   $^{90}$Zr and  $^{132}$Sn, the main GT peak is shifted 500keV lower by the EDF SAMi-T with tensor terms, but slightly higher in energy for the small shoulder peak.   The quenching factor is $qf=0.65$ and $qf=0.66$ for SAMi and SAMi-T, respectively.

Experimental Gamow-Teller (GT)  decay matrix elements were studied in different mass region in comparisons with shell model calculations.
Experimental Gamow-Teller strengths for $p$-shell with mass $A<16$ \cite{137} requires an effective Gamow-Teller operator $\sigma\tau$ multiplied by 0.82$\pm$0.015,  or
equivalently,  the axial vector coupling $g_A=1.27$ is replaced by an effective value $g_A(eff) =(0.82\pm0.015)g_A$.
It should be noticed that our $qf$ is squared of the quenching factor introduced for the  Gamow-Teller operator; $qf=0.82^2=0.67$ for the
GT decay probability.
A similar phenomenological correction, $g_A^{eff} = (0.77\pm0.02)g_A$
brings shell model predictions into agreement with data for $sd$-shell nuclei with $16<A<40$ \cite{199}.
 For $pf$-shell nuclei
with mass number $A$ between 41 and 50, shell model results  are compared with the GT beta decays,
 in ref. \cite{139},
and found most of the experimental data quite
well reproduces with the effective Gamow-Teller operator $\sigma\tau$ multiplied by 0.744$\pm0.015 (qf=0.554\pm.022)$ .
Similar studies with difference effective interactions are performed in ref. \cite{200}  for $pf-$shell and $f_{5/2}pg_{9/2}$ configurations for mass $Z=20-30$ and $N<50$ and a slightly larger  correction $g_A^{eff} = (0.660\pm0.016)g_A$ and  $g_A^{eff} = (0.684\pm0.015)g_A$ are preferred for  neutron-rich $pf$-shell nuclei, $50<A<70$, and   for $f_{5/2}pg_{9/2}$-shell nuclei, $70<A<80$ regions, respectively. 
Such a simple renormalization can provide an effective prescription to cure
 a theoretical problem
that has been discussed  for several decades \cite{146, 201,203,204,205}, although the quenching factor varies from 0.660 to 0.744  in $pf$-shell nuclei depending on
the adopted interactions.

In general, the agreement between RPA results and experimental data of GT giant resonances are satisfactory.  Especially the SAMi-T gives
better agreement for the small peak below GTR.  This is due to a fine tuning of the spin-orbit splitting near the Fermi energy by the tensor terms in EDF \cite{SAMi-T}.

\subsection{SD strength}

\begin{table*}
\caption{Sum rules of GT$_-$, SD$_-$ and SD$_+$ resonances for $^{48}$Ca, $^{90}$Zr, $^{132}$Sn and $^{208}$Pb. A quenching factor is extracted by comparing the experimental and theoretical sum rules. The theoretical results are calculated by using SAMi (SAMi-T) EDF. The unit of sum rule values for SD$_-$ and SD$_+$ is fm$^2$.  }
\begin{ruledtabular}
\begin{tabular}{cccccccc}
   & && GT &  & & SD &   \\
   Nuclides & channel& Exp. &  Theo. & $qf$ & Exp. & Theo. &$qf$       \\

\hline
 $^{48}$Ca   & ($t_-$)& 15.3 $\pm$ 2.2  & 23.96 (23.74)   &  0.64 (0.64)  & 97  $\pm$ 11 (total)          & 148.69 (133.43) & 0.66 (0.73)  \\
 $^{90}$Zr  & ($t_-$) & 20.75          & 29.89 (29.59)   &  0.69 (0.70)  & 247 $\pm$ 20(total)  & 314.27 (305.05) & 0.78 (0.81)  \\
 $^{90}$Zr   & ($t_+$)&                 &                 &               &  98 $\pm$ 9 (total)  & 169.49 (168.72) & 0.58 (0.58) \\
 $^{132}$Sn  & ($t_-$)& 53 $\pm$ 15 ($t_-$)    & 95.61 (93.86)   &  0.55 (0.56)  &                              &                 &             \\
 $^{208}$Pb & ($t_-$) & 85.00 ($t_-$)          & 130.65 (128.05) &  0.65 (0.66)  & 1004 $\pm$ 23(total) & 1279.6 (1256.2) & 0.78 (0.80)   \\
 $^{208}$Pb  (0$^-$) & ($t_-$) &                 &                 &               & 107  $\pm$ 7& 169.86 (159.34) & 0.63 (0.67)   \\
 $^{208}$Pb (1$^-$) & ($t_-$) &                 &                 &               & 450  $\pm$ 15& 443.03 (439.71) & 1.02 (1.02)   \\
 $^{208}$Pb (2$^-$)  & ($t_-$)&                 &                 &               & 447  $\pm$ 15& 666.71 (657.16) & 0.67 (0.68)   \\
\end{tabular}
\end{ruledtabular} \label{sum-gt-sd}
\end{table*}

\begin{table*}
\caption{The total RPA non-energy weighted sum rules m$_0$(t$_-$), m$_0$(t$_+$), and m$_0$(t$_-$)-m$_0$(t$_+$) of SD resonances for $^{48}$Ca, $^{90}$Zr, $^{132}$Sn and $^{208}$Pb.  The sum rule values of m$_0$(t$_-$)-m$_0$(t$_+$) from the analytic formula \eqref{eq:SD-sum} are also presented in the table. The  values of $^{208}$Pb for the J$^{\pi}$= 0$^-$, 1$^-$, and 2$^-$ components are shown in the last three lines. The results are calculated by using SAMi (SAMi-T) EDF. The unit of sum rule values is fm$^2$. }
\begin{ruledtabular}
\begin{tabular}{ccccccccc}
            & SAMi   &       &            &      & SAMi-T &       &            &       \\
   Nuclides & $m_0(t_-$)  & $m_0(t_+$) &  $m_0(t_-)-m_0(t_+$) & $m_0(t_-)-m_0(t_+$) & $m_0(t_-)$  & $m_0(t_+$) &  $m_0(t_-)-m_0(t_+$) & $m_0(t_-)-m_0(t_+$)    \\
             &         &        &         &  from eq.(5) &     &        &         & from eq.(5) \\
\hline
 $^{48}$Ca          & 159.59  & 67.17  & 92.42   &  92.41   & 157.67  & 67.19  & 90.48   & 90.46   \\
 $^{90}$Zr          & 323.33  & 173.45 & 149.77  &  149.78  & 320.12  & 173.52 & 146.40  & 146.60  \\
 $^{132}$Sn         & 740.52  & 120.90 & 619.62  &  620.03  & 728.66  & 113.46 & 615.21  & 615.23  \\
 $^{208}$Pb         & 1281.53 & 194.86 & 1086.67 &  1086.68 & 1260.13 & 182.72 & 1077.41 & 1077.45 \\
 $^{208}$Pb(0$^-$)  & 170.31  & 49.56  & 120.75  &  120.74  & 159.99  & 40.27  & 119.73  & 119.72  \\
 $^{208}$Pb(1$^-$)  & 443.63  & 81.40  & 362.23  &  362.23  & 440.15  & 80.99  & 359.16  & 359.15  \\
 $^{208}$Pb(2$^-$)  & 667.59  & 63.90  & 603.66  &  603.71  & 659.99  & 61.46  & 598.52  & 598.58 \\
\end{tabular}
\end{ruledtabular} \label{tab:SD-sum}
\end{table*}

The calculated SD strength for   $^{48}$Ca, $^{90}$Zr  and  $^{208}$Pb are shown in Figs. \ref{SD-Ca}, \ref{SD-Zr}, \ref{SD-Pb}, respectively.
Results of multipole decomposition with $J^\pi$=0$^-,1^-$ and 2$^-$ in $^{48}$Ca and  $^{208}$Pb are shown in  Figs. \ref{SD-Ca-J} and  \ref{SD-Pb}, respectively.
The tensor interactions have substantial  effects on SD response, especially the effect is different for each multipole.   For 0$^-$ and 2$^-$, the tensor effect makes the strength distributions higher in energy, while $1^-$ response is shifted lower in energy as is shown in Figs.  \ref{SD-Ca-J}, \ref{SD-Pb}.  As a net effect, the main peak  at E$_x\sim$23MeV is shifted 1MeV lower in energy by the tensor effect and give a better description of the experimental strength distributions  of SD for $^{48}$Ca.  The same trend is also found for $t_-$ channel of SD strength for $^{90}$Zr and show a fine agreement with the experimental data.  The $t_+$ channel of SD strength from  $^{90}$Zr  is also shown in
Fig.  \ref{SD-Zr}.  Calculated strength is distributed in the energy region E$_x$=0$\to$20MeV, while experimental data show even some strength in a region between 20-35MeV.  For $^{208}$Pb,  the strength for $J^\pi=0^-$ is shifted about 8MeV upward in energy, while that for $J^\pi=1^-$ is shifted 4MeV downward by the tensor effect.  This is already noticed in ref. \cite{21} as the hardening and the softening effect by the tensor interactions on SD strength.  The  $J^\pi=2^-$  response gets also a hardening effect, but smaller than that for $J^\pi=0^-$.  The empirical summed strength in the top panel of Fig. \ref{SD-Pb} is better described by SAMi-T EDF than SAMi  EDF without the tensor terms.

The quenching effect is modest in general for SD strength.  In $^{48}$Ca, the $qf$ value is 0.64 for  the summed GT strength, but $qf=0.66 (0.73)$ for the SD strength with SAMi (SAMi-T) EDF.   In the case of  $^{90}$Zr, the GT  strength needs the $qf=0.7$, while the SD strength
shows $qf=0.8$.  In $^{90}$Zr also, the tensor effect makes a slightly modest $qf$-value with 0.81 for SAMi-T and 0.78 for SAMi.  The feature of quenching is the same also for $^{208}$Pb; $qf=0.65$ for GT strength and $qf=0.78(0.80)$ for SD with SAMi (SAMi-T) EDF.
The obtained quenching factors are summarized in Table I.

The sum rule values of RPA calculations and also of analytic equation \eqref{eq:SD-sum} are tabulated  in Table \ref{tab:SD-sum}.  There is a small difference in the RPA sum rule values of SD $t_-$ strengths in Tables \ref{sum-gt-sd} and  \ref{tab:SD-sum} since all strengths are accumulated in Table  \ref{tab:SD-sum}, while the maximum energy for accumulation is taken to be the same as the experimental maximum energy  in Table  \ref{sum-gt-sd}.  One can see from Table \ref{tab:SD-sum} that the $t_+$ channel has appreciable strength, about 40\% in $^{48}$Ca and 15\% in $^{208}$Pb of the values of the $t_-$ channel,  since the SD states are 1$\hbar \omega$  particle-hole excitations and
the neutron excess does not block completely the $t_+$ channel.  In the case of the GT sum rule, the blocking of neutron excess prohibits the $t_+$ excitations completely and the $t_-$ sum rule value exhausts almost all the Ikeda sum rule strength,  $3(N-Z)$,  as is seen in Table. \ref{sum-gt-sd}.
The proportionality $(2\lambda+1)$ holds precisely for the sum rule values
$m_0^\lambda(t_-)-m_0^\lambda(t_+)$ in $^{208}$Pb.  On the other hand, this proportionality does not hold for $m_0^\lambda(t_-)$ value since $m_0^\lambda(t_+)$ has some irregularities and becomes  the largest for $1^-$ excitations.  In general the tensor effect on the sum rule values  is small for both $m_0(t_-)-m_0(t_+)$  and $m_0(t_-)$ values about $1-2$\% effect, except for the
$m_0(t_-)$ value for the 0$^-$ case in $^{208}$Pb.

\section{Summary}
We studied the GT and SD strength distributions of doubly closed shell nuclei  $^{48}$Ca,  $^{90}$Zr,   $^{132}$Sn, and $^{208}$Pb
with a self-consistent HF+RPA method with Skyrme-type EDFs,  SAMi and SAMi-T.  In the latter, the tensor terms are included by means of  the "ab initio" type model based of AV18 interactions.  The gross features of both GT and SD strength distributions are well  reproduced by the
present calculations.  Especially the main peak positions of both resonances are  described well by the calculated results.
The tensor interactions have a small effect on GT states, but the small low energy GT peaks of $^{48}$Ca,  $^{90}$Zr,   and $^{132}$Sn are better described by the EDF SAMI-T with the tensor terms.  For SD response, the tensor effect is much larger and different for each multipole; the hardening effect of $J^\pi=0^-, 2^-$ peaks and the softening effect on  the $J^\pi=1^-$ peak.  The accumulated strength is larger in the case of SAMi-T than that of SAMi up to E$_x$=30MeV.
In general, the quenching effect is modest for SD strength with the quenching factor  $qf\sim0.8$ compared with that for GT,   $qf\sim(0.55-0.69)$, which is consistent with the quenching value obtained from the GT beta decay processes in nuclei $A<50$.   This difference in the effective quenching factors between GT and SD should be implemented in future theoretical study of double beta decay probabilities.  It will also be a future project to study this difference with microscopic models beyond RPA as well as with two-body currents.

\section*{ACKNOWLEDGEMENTS}
We would like to thank K. Yako for valuable discussions and also proving us with a preliminary version of the SD strength of $^{48}$Ca.
We would like to thank also T. Uesaka, T. Wakasa, M. Sasano and T. Suzuki for fruitful discussions.
This work was supported  by JSPS
KAKENHI  Grant Number  JP19K03858 and by the National Natural Science Foundation of
China under Grant Nos 11575060, 11775014 and 11975096.

\end{document}